# A Model for an Intelligent and Adaptive Tutor based on Web by Jackson's Learning Styles Profiler and Expert Systems

H. Movafegh Ghadirli and M. Rastgarpour

*Abstract*—Todays, Intelligent and web-based E-Learning is one of the important area in E-Learning. This paper integrates an intelligent and web-based E-Learning with expert system technology in order to model the learning styles using Jackson's model. It is intelligent because it can interact with the learners and offer them some subjects based on Pedagogy view. Learning process of this system is in the following. First system determines learner's individual characteristics and his learning style based on a questionnaire in Jackson's learning styles profiler. Learning styles profiler is a modern measure of individual differences in learning style. Then learner's model is obtained and an Expert system simulator plans a *"pre-test"* and then rates him. The concept would be presented if the learner scores enough. Subsequently, the system evaluates him by a *"post-test"*. Finally the learner's model would be updated by the modeler based on try-and-error. The proposed system can be available Every Time and Every Where (ETEW) through the web. It has all good facilities such as hypertext component, adaptive sequencing, problem solving support, intelligent solution analysis and adaptive presentation. It improves the learning performance and has some important advantages such as high speed, simplicity of learning and low cost.

## I. INTRODUCTION

From 1980, the computers have been applied to develop electronic learning (E-Learning) systems. Many researches have been done in this field to improve its efficiency so far. E-learning can be defined as an activity to support a learning experience by either developing or applying information and communication technology (ICT) [1].

The first E-Learning softwares were using some Medias such as CDs or Web before. They are static, non-intelligent and inflexible. Because a course had been organized by prior procedure and taught in the same style for all learners. They were either computerized versions of textbooks (characterized as electronic page turners) or drill and practice monitors that presented a learner with problems and compared learner's responses to pre-scored answers.

Learners are different. It leaded to decrease the efficiency of this style. It means that some learners need to repeat some lessons meanwhile some lessons must be removed for others. Later the researchers concluded that learning process must be dynamic and intelligent based on pedagogy view [2]. But developing an applicable and trustful system is so hard [3]. This caused to advent new generation of intelligent educational systems [4].

A human expert tutor can adapt a proper sequence of lessons and training speed in according to learner's aptitude and characteristics. He can apply some educational tools and adjust the diction based on learner. He sometimes cancels the class when learner is not in proper situation. But human expert tutor is infrequent and expensive. This is another reason to need to new generation of intelligent educational systems.

Nowadays, "web-based learning" and "intelligent learning" is one of the most significant topics in education [3], [5]. A web-based tutor has some benefits like tirelessly, dominance on concepts, low cost and invariant of time and place. It can utilizes conversational agents as well [6]. However millions learners of the world can learn by thousands of expert tutor through the web in an intelligent and virtual schools. One of the important classes of intelligent tutors is based on rule-based expert system. It determines whatever the student knows, doesn't know and knows incorrectly, and uses this information to adjust learning style. This education type which derives the benefits of Expert Systems is called *model tracing tutor* [4].

This paper introduces an intelligent system to enjoy Expert Systems abilities. So E-Learning would be efficient, adaptive and just needs a computer and internet connection. Adapting with web-based contexts is very important, because the concept, which is developed for one user, isn't useful for others [3].

The proposed system determines the learning style by a test. So it develops a primary model of learner. The learning process starts then. Gradually, some characteristics of learner may be varying during learning progress. These improvements would be saved by system. So learner model gets more accurate step by step. System can receives scientific and mental feedback of learner and then change the learning style during the process.

Manuscript received January 07, 2012; revised January 27, 2012.

Hossein Movafegh Ghadirli is a graduate Student in Computer Engineering and with Young Researchers Club, Islamshahr Branch, Islamic Azad University, Islamshahr, Iran. (email: hossein.movafegh@iau-saveh.ac.ir ; hossein.movafegh@gmail.com )

Maryam Rastgarpour is an instructor with the department of Computer Engineering, Saveh branch, Islamic Azad University, Saveh, Iran.
(E-mail:m.rastgarpour@iau-saveh.ac.ir;m.rastgarpour@gmail.com)

Web-based learning is useful not only for high level education environments such as universities but also for training new employees and adapting staffs according to company changes. Ideally, web-based learning improves the learning efficiency for students and employees through features that are not available in face to face learning. It also allows learners to access the learning materials and interact with the rest of course ETEW. The aim of proposed system is to offer the content which the learner can't be aware of it easily.

The rest of this paper is organized as follows. Section II defines an intelligent E-Learning system and presents some available samples. It also deliberates adaptive E-Learning, Jackson Model and learning styles. Section III describes the proposed E-Learning system which is intelligent, adaptive, customized and web-based. Finally this paper concludes in Section IV.

## II. BACKGROUND

### A. Intelligent E-Learning system

The first intelligent E-Learning systems were computer-aided instruction (CAI) systems which presented the early 1960's. Furthermore the adaptive intelligent systems are not novel at all. All of these systems are a kind of "Intelligent Tutoring Systems (ITS)" or "Adaptive Hypermedia Systems" [4].

The late 1960s, ITSs have moved out of the lab and into classrooms and workplaces where some have been shown to be highly effective [7], [8]. While intelligent tutors are becoming more common and proving to be increasingly effective, they are difficult and expensive to build.

E-Learning systems are classified into two classes: *intelligent* and *non-intelligent*. Non-intelligent systems apply the same style for variant learners. It is the worst disadvantage of them, because the learners are various in terms of awareness and mentality. So it intensely needs to organize course contents intelligently.

Intelligent systems realize the customized and adaptive E-Learning based on course content, learner type and education method [4]. This system can recognize the student type, choose appropriate course content from knowledge base and present it to learners in proper style. It also attempts to simulate a human tutor and act expertly and intelligently. Students using these systems usually solve problems and related sub-problems within a goal space, and receive feedback when their behavior diverges from that of the learner model. Figure 1 shows the process of intelligent E-Learning system.

Some available intelligent E-Learning systems are introduced in Table I to handle the pragmatics of three elements: content, learner model and education methods for adaptive and customized learning. These systems have been validated to simplify learning.

### B. Adaptive E-Learning

The adaptive E-Learning has an important role in efficiency of educational environments. These systems can also be compatibles with heterogeneous population of learners. They could defeat similar version of non-adaptive systems, because an efficient, adaptive and dynamic E-Learning can recognize learning style of learner based on pedagogy principles. It can adapt the learners with current status of system. Then it changes its behavior dynamically and presents the learning concepts according to learner's model.

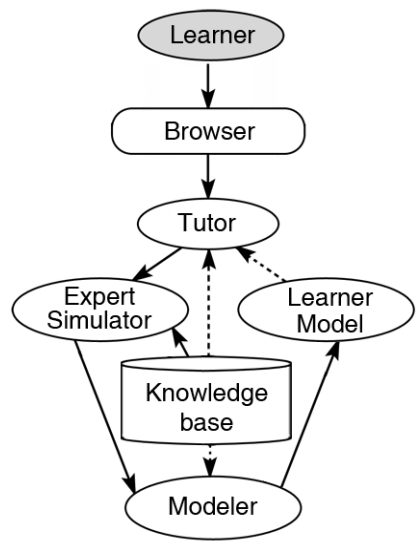

Fig. 1. Intelligent E-Learning System

*1) Jackson's Learning Styles Profiler*

This section introduces Jackson model [9]. There are some approaches to model learner's behavior. For example Entistle approach [10] tries to make connections between psychology concepts and effective variables on learner's view to *'learning'*. Despite they have some benefits in education and psychology of personality; they are rarely applied in adaptive education systems.

Jackson's model is based on the most recent researches in psychology of personality. It also is efficient analyzer. So this paper just applies it among available models.

Jackson model has been proposed in Queensland University [9]. He has investigated personality, learning and evaluation for fifteen years. Learning is based on new neurological psychology in this model.

This model has some advantages. For example, it can report a learner individually. It is obtained on basis of the

TABLE I
SOME AVAILABLE INTELLIGENT E-LEARNING SYSTEMS

| Name | Comments |
|---|---|
| *CAES system* [11] | Developed by integration of shipping simulation and intelligent decision system. The task is to teach shipping to captains in virtual turbulent sea. |
| *ICATS* [12] | Coordinates an expert system with multimedia system in an intelligent learning system. |
| *ANDES* [13] | Trains Physics without using natural language. |
| *ATLAS* [14] | Uses natural language to teach Physics. It can reply to learner's questions. |
| *ELM-ART* [15] | Web-based tutoring System for learning programming in LISP. |
| *APHID-2* [16] | A hypermedia generation system with adaption by defining rules that map learner information. |
| *AUTOTUTOR* [17] | Be full intelligent and teach *"introduction to computers"* in at least 200 universities of the |

points which the learner earns in learning style ( section III-

B).

The aim of this model is to recognize an ideal style based on Pedagogy principles which can model learner's behavior and specifications in adaptive E-Learning system. A key premise of Jackson's model is that cognitive strategies redirect Sensation Seeking to predict functional behaviors.

*2) Learning Styles*

Learning styles are various approaches or methods of learning. An adaptive E-Learning system is based on accurate recognition of behaviors and individual characteristics. There are some important factors in learning style such as aptitude, personality and behavior [2]. It's worth to be noted the concept of "learning style" has emerged as a focal point of much psychological research. In our proposed learner model, we have adopted the Jackson's learning styles profiler in order to model learning styles of learners. Jackson proposes five learning styles which is summarizes in Table II.

**Sensation Seeking (SS):** These people are *impulsive* and *hurried*. New situations are exciting for them so that they can't wait and would like to experience and explore it immediately. They believe to "*action*" and can perform multiple tasks simultaneously. These people would rather to explore their environment by themselves and also learn by test and error.

**Goal Oriented Achievers (GOA):** They adjust certain and difficult aims. They try to increase their abilities by skills and required cognitive resources in order to realize their aims. They think troubles are instructive challenges. Furthermore they believe that can realize to whatever they want.

**Emotionally Intelligent Achievers (EIA):** Emotional independence and rational thinking are their noticeable characteristics. They are patient learners and can have the best efficiency after perceiving logic of problem. They can generalize well and often divide a problem to small and understandable parts to solve them.

**Conscientious Achievers (CA):** They are responsible and clever. They can learn well by collecting, analysis and review some information before action. They prefer to analyze all aspects of a problem. Thus they can connect discrete data to each other and avoid a mistake. These people usually have extensive knowledge in area of interest.

**Deep Learning Achievers (DLA):** they have deep perception of concepts. They want to know how can use previous knowledge in practice. They can learn well when they know practical value of something. Thus they can test that theory or idea. In fact, learning is difficult for them, when they don't know the target [2].

TABLE II
SUMMARIZATION OF LEARNING STYLES

| Learning Style | Specifications |
|---|---|
| SS | Believing experiences create learning. |
| GOA | Self-confident to achieve difficult and certain target. |
| EIA | Rational and goal-oriented. |
| CA | Responsible and insight creator. |
| DLA | Interested in learning highly. |

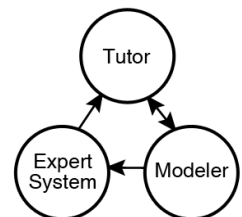

Fig. 2. Intelligent Tutor Elements

### III. PROPOSED APPROACH

A web-based, adaptive and intelligent tutor is an E-Learning system based on web which can be used ETEW. It can determine learner style, learning content and presentation technique adaptively. It is updated automatically with learner's characteristics and behavior. The first E-Learning system which is web-based and intelligent, has been reported on 1995 [4], [18]. Learning of all courses is customized well at home via web in this system. So learners can solve some examples and proper exercises ETEW. Finally he can take a course exam which would be virtual or physical. Figure 2 show the elements of intelligent tutor.

*A. Learning Environment*

Learner can visits website of intelligent virtual school after authentication. An intelligent Graphical User Interface (GUI) is an interface between learners and intelligent tutor. This section of system can affect learning efficiency. It should be user friendly.

An intelligent virtual class has some properties such as graphical properties, audio and video to make learning attractive. Moreover, some tools are available to simplify learning process. Learners can communicate well with this inanimate and non-physical system by these tools. Some facilities are Computer games, frequently asked questions (FAQ), Video chat and email.

*B. Training Method*

Knowledge of expert tutor includes of two parts, learning content and learning technique. Learning content is theoretic information, technical content and probably experiments which expert tutor does. Learning technique is some experiences which he have got during teaching years [19]. An expert tutor teaches educational content in proper method by film, dynamic view, game and even puzzle. He gets feedback from learner during training. So learner level may be changed than start learning. Tutor helps learner by "*the best way*" in proposed model.

The expert tutor offers an education method based on learner's type. Furthermore each course section has individual significance which is different from others. Tutor often determine different scores for variant sections according to education method. Moreover he rates highest one to the most important section of course in all education methods.

The smallest part of any topic which can't be divided more is called "concept". It is usually equal to a lesson in physical class. Educational concepts would be transferred to knowledge base in this system. Then the system can distinguish all concepts and relocate all parts.

Sometimes a lesson is needed to repeat, relocate or even remove for a learner. Most of available systems intelligently guide a learner to a special aim while only a few intelligent systems allow leaner to select subsections of a concept.

This system uses a three layered structure to implement a concept: *"Pre-test"*, *"Learning concept"* and *"Post-test"*.

The pre-test includes of some questions planned by an expert tutor. These questions can determine learner's primary knowledge level. It simulations learner model based on Jackson model [9]. The learning concept depends on learner level. So the best method to train a learner is determined. Afterwards learning process starts up. Subsequently a post-test evaluate the learner by some questions. Figure 3 shows block diagram of proposed system.

In proposed system, the teacher can react in two ways: by sending a message either to the learner model or directly to the learner via e-mail. Sending a message to the learner model is very simple and works even if the learner did not provide an e-mail address.

*C. Learning Evaluation*

Learning evaluation is the main factor to determine learning performance in E-Learning systems. The evaluation has two levels, *conceptual* and *objective*. The first one refers to learner's understanding of the lesson concept. The last one denotes to learner's understanding of the lesson topic. Learner's knowledge level is determined by concept level and objective level.

The tutor can extract proper questions from question base by an expert system, pre-test and post-test. He notes that a specific score is given to each question.

Question selection should satisfy some rules. Firstly none of them should be repetitious even if a learner would be trained one concept several times. Secondly the question must be planned for all sections of a concept entirely. Thirdly expert tutor plans questions in all level.

TABLE III
CATEGORIES OF KNOWLEDGE LEVEL

| Knowledge Level | Score |
|---|---|
| *Excellent* | 86-100 |
| *Very good* | 71-85 |
| *Good* | 51-70 |
| *Average* | 31-50 |

Sequence, number and level of questions are determined according to learner level and learning type intelligently. Sum of scores is calculated and learner level is determined after answering the questions. Table III presents five categories of learner's knowledge level about a concept [20], [21].

The modeler of system updates the learner's model during progress of answering the questions. It can also save last academic status of learner and all his learning records.

Table IV summarizes the proposed system and some available ones in terms of different properties like hypertext component, adaptive sequencing, problem solving support, intelligent solution analysis and adaptive presentation [22]. It's clear in this table that available systems don't have all of these useful properties while proposed system has been designed to covers all.

IV. CONCLUSION

This paper proposed a model for an adaptive, intelligent and web-based tutor which is able to model the learning styles of learners using Jackson's learning styles profiler and expert system technology to improve learning performance. Previous E-Learning systems offer fixed multimedia web pages and facilities for user management and communication.

Based on the proposed model, E-Learning software is implemented on the web which can identify learning styles, aptitude, characteristics and behaviors of learner in order to provide appropriate individual learning content.

Jackson model can recognizes learner's learning style. It leads to make learner's model accurately. So the efficiency of adaptation process increases.

The proposed model acts as an intelligent tutor which can perform three processes - *pre-test*, *learning concept* and

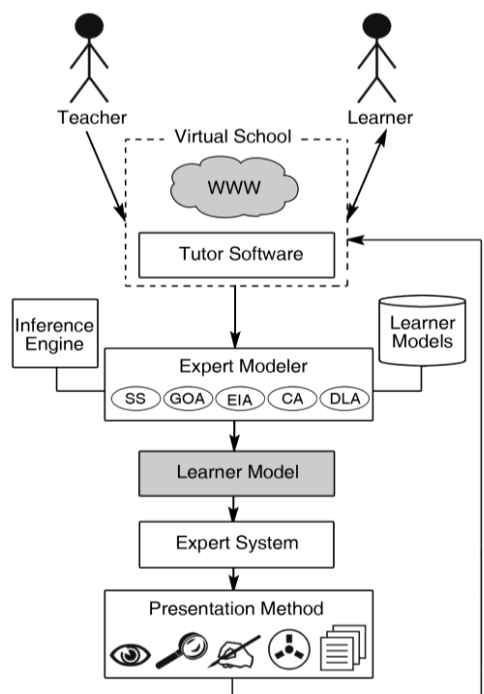

Fig. 3. Web-based Education System

TABLE IV
ADAPTATION TECHNOLOGIES IN WEB-BASED SYSTEMS

| System | Hypertext Component | Adaptive Sequencing | Problem Solving Support | Intelligent solution Analysis | Adaptive Presentation |
|---|---|---|---|---|---|
| AST | Y | Y | | | Some |
| *InterBook* | Y | Y | | | Some |
| *PAT-InterBook* | Y | Y | Server | Y | Some |
| *DCG* | Y | Y | | | |
| *PAT* | Y | | | | |
| *WITS* | N | | | Y | |
| *C-Book* | Y | | | | Y |
| *Manic* | Y | Some | | | |
| *Proposed System* | Y | Some | Server | Y | Y |

*post-test* – based on learner's characteristic. This system uses expert simulator and its knowledge base as well.

It is also web-based which leads to be low-cost and available everywhere and every time. Consequently thousands of students can learn simultaneous and integrated efficiently.

The proposed system cover all important properties such as hypertext component, adaptive sequencing, problem solving support, intelligent solution analysis and adaptive presentation while available systems have only some of them. It also doesn't have any drawback of previous system and human expert tutor. It can significantly improve the learning result. In other words, it helps learners to study in "*the best way*".

## V. ACKNOWLEDGEMENT

The first author would thank Dr. F. Taghiyareh so much for her useful guidance during his studying.